\documentclass[journal]{IEEEtran}
\usepackage[pdftex]{graphicx}
\usepackage{amsmath}
\usepackage{amssymb}
\usepackage{blindtext}
\usepackage{color}
\usepackage{epstopdf}
\usepackage[pdfstartview=FitR]{hyperref}

\ifCLASSINFOpdf
\else
\fi

\begin{document}
%
\title{Momentum-Space Topological Effects of Nonreciprocity}


\author{\IEEEauthorblockN{S. Ali Hassani Gangaraj,~\IEEEmembership{Student member,~IEEE} and George W. Hanson,~\IEEEmembership{Fellow,~IEEE}}  \\\ \large{(Invited Paper)}

\thanks{Manuscript received xxx; revised xxx.

G. W. Hanson is with the Department of Electrical Engineering, University of Wisconsin-Milwaukee, Milwaukee, Wisconsin 53211, USA (e-mail: george@uwm.edu).

S. Ali Hassani Gangaraj is with the Department of Electrical and Computer Engineering, Cornell University, Ithaca, New York 14583, USA (e-mail: ali.gangaraj@gmail.com ).

}}

\markboth{Journal of \LaTeX\ Class Files,~Vol.~xxx, No.~xxx, xxx~2015}%
{Shell \MakeLowercase{\textit{et al.}}: Bare Demo of IEEEtran.cls for Journals}
%


\maketitle

\begin{abstract}
The connection between topology and nonreciprocity in photonic systems is reviewed. Topological properties such as Chern number, and momentum-space properties such as Berry phase and Berry connection, are used to explain back-scattering immune edge states and their topological protection. We consider several examples to illustrate the role of momentum-space topology on wave propagation, and discus recent magnet-less approaches.
\end{abstract}

\begin{IEEEkeywords}
Berry Phase, Berry curvature, photonic topological insulator.
\end{IEEEkeywords}

\maketitle

\section{Introduction}\label{Intro}

\IEEEPARstart{I}{}n this work, we review how topology affects, and can be used to understand, novel electromagnetic phenomena associated with nonreciprocity. Of particular interest are topologically-protected (TP) edge modes, immune to backscattering and diffraction. We consider systems that break time-reversal (TR) symmetry, which are the photonic analogs of the quantum Hall effect; alternatively, in reciprocal systems, breaking inversion (I) symmetry leads to photonic analogs of quantum spin Hall insulators \cite{Rev1}-\cite{Rev03}.

Various physical quantities reverse their direction under time reversal (e.g., magnetic field and magnetic vector potential, particle velocity, and hence particle kinetic momentum and current density), such that applying a DC current or magnetic field bias can be used to break TR. Breaking TR symmetry via gyrotopic effects in biased ferrites or plasmas \cite{Mario2}-\cite{Hassani3} necessitate a strong static magnetic field bias, which leads to severe restrictions in size, weight, cost, and integrability. Magnet-less approaches to break TR, which are the focus of this special issue, include time modulation or mechanical motion \cite{Fan}-\cite{Hadad}, applying a DC bias current \cite{D1}-\cite{D2}, metamaterials integrated with transistors \cite{Caloz_1}-\cite{Caloz_4}, and material non-linearities \cite{NL1}-\cite{Engheta_1}. We can also mention novel natural systems where nonreciprocity is produced, such as topologically-protected ocean and atmospheric equatorial edge modes that propagate energy along the Earth’s equator; in this case, TR symmetry is broken by the earth's rotation \cite{TOEW}. 

Transistor-based nonreciprocity leads to a magnet-less approach, requiring, rather than a DC magnetic bias, a simpler DC voltage bias for the field-effect transistors (FETs) \cite{Caloz_1}-\cite{Caloz_4}. Various transistor configurations can be used to, e.g., emulate a magnetic moment that mimics the rotating magnetic moment in ferrites due to electron spin precession. Such metamaterials may behave as a photonic crystal, or as an effective medium, depending on the structure and wavelengths involved. 

Another method to design a non-reciprocal structure without magnetic bias is via nonlinearities \cite{NL1}-\cite{Engheta_1}. For example, in \cite{Engheta_1}, a metasurface has been proposed where the nonreciprocity is achieved by incorporating nonlinear (varactor-loaded), resonant structures inside an asymmetric one-dimensional dielectric slab.

\section{Topology, Trivial and Non-Trivial Bandgaps}

A key ingredient to obtain a TP edge mode is to have degenerate bulk dispersion bands that open a \textit{non-trivial} band gap when TR symmetry is broken, resulting in a gap-crossing edge mode. Opening or closing a bandgap is a change of the topology of the bulk momentum-space dispersion surface, which, surprisingly, governs the propagation of TP edge modes at material interfaces via bulk-edge correspondence.

An invariant that has primary importance in topology is the genus $g$ of a surface, essentially, the number of holes that it has. The Gauss-Bonnet theorem \cite{Frankel} states that for a closed surface, the integral of the Gaussian curvature $K$ over the surface is related to the genus as 
\begin{equation}
\oint_{S}K dS =2\pi(2-2g). \label{GB}
\end{equation}
A sphere has genus 0 (no holes, $K=1/R^2$), and so $\oint_{S}K dS =4\pi$, the solid angle of the sphere, whereas a toroid has genus 1 (one hole), and so $\oint_{S}K dS =0$. Any two objects having the same genus can be continuously deformed one to the other, such as a toroid deforming into a teacup. 

Alternatively, for dispersion surfaces in momentum space, a quantity known as the Berry curvature $\mathbf{F}$ replaces the geometric curvature $K$, and a quantity known as the Chern number $C_n$ replaces the genus, and satisfies 
\begin{equation}
\oint_{S}\mathbf{F}\cdot \boldsymbol{\mathrm{n}} \,dS =2\pi C_n \label{CN}
\end{equation}
as detailed later, where $S$ is the momentum-space Brillouin zone surface for periodic media (equivalent to a toroid), and to the Riemann sphere for continuum media  \cite{Mario2}). The Chern number, like the genus, is a topological invariant, i.e., it can only change upon a change of topology.

\begin{figure}[ht]
\begin{center}
\noindent  \includegraphics[width=3.5in]{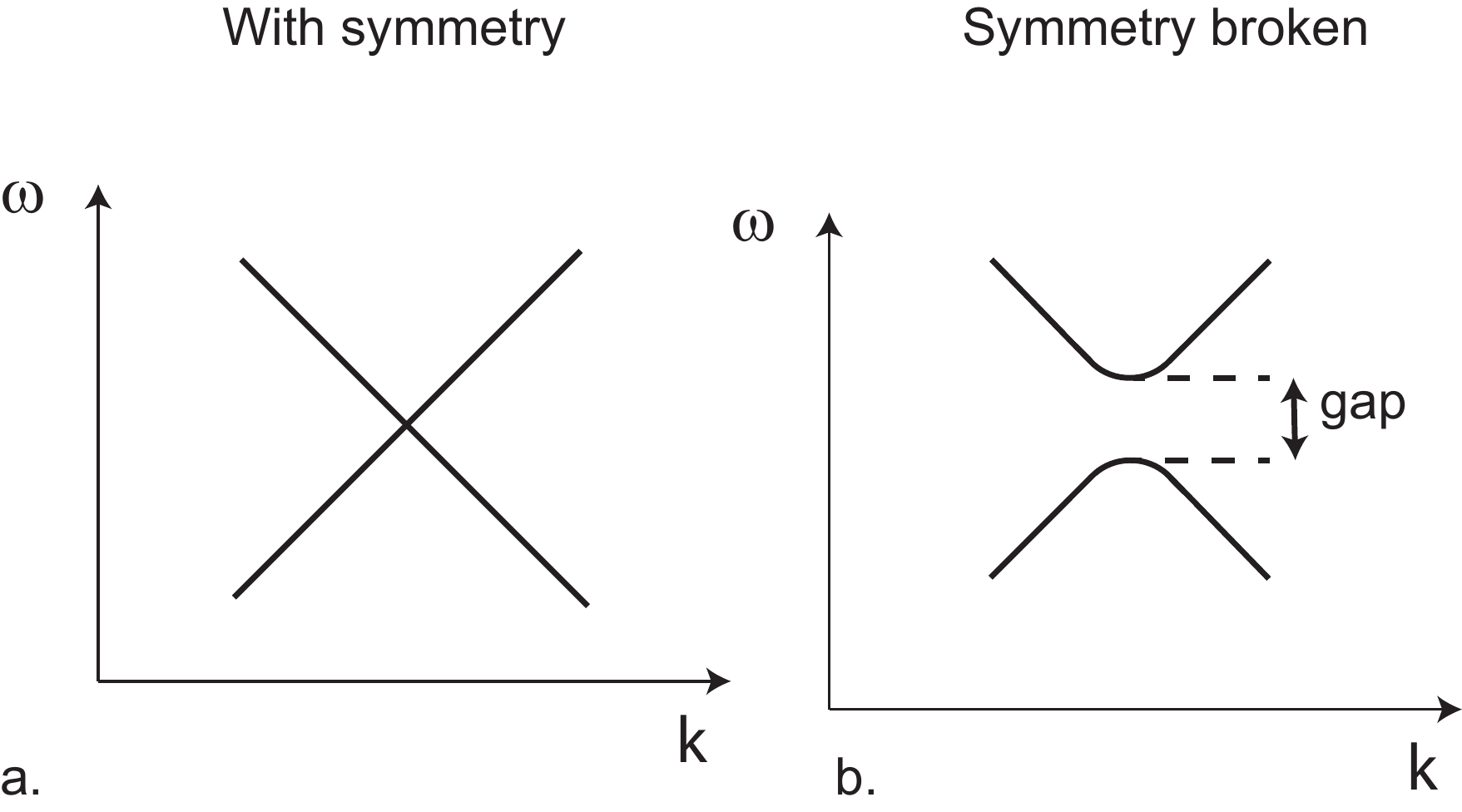}  
\end{center}
\caption{(a) Bulk mode dispersion in the presence of TR symmetry leading to a band degeneracy. b. TR symmetry is broken, lifting the degeneracy and opening a non-trivial bandgap.}
\label{BS}
\end{figure}

Regardless of the source of nonreciprocity, the crux of understanding the effect of momentum-space topology on photonic systems in a physical sense is the opening of a \textit{non-trivial} band gap in (i.e., changing the topology of) the bulk material dispersion when TR symmetry is broken, as depicted in Fig. \ref{BS}. 
The non-trivial nature of the opened gap is important. When a non-trivial bandgap is opened (e.g., a periodic array of nonreciprocal  cylinders immersed in a dielectric host, or alternating layers of nonreciprocal dielectric slabs), the eigenmodes have a sort of rotation as a function of momentum (quantified below as the curl of the Berry connection), leading to a non-zero Chern number \cite{Rev1}. In a simple medium, when the opened gap is trivial, the eigenmodes don't have the required rotation as a function of momentum. 

\section{Berry Phase Arising from Hermitian Eigenvalue Problems in Momentum Space}

From a mathematical view, the important concept to understand nontrivial topologies is to appreciate the gauge ambiguity of the Hermitian eigenproblem. Consider a general Hermitian eigenvalue problem, typically arising from the
spatial Fourier transform $\mathbf{r}\leftrightarrow \mathbf{k}$ of a
space-domain wave equation (e.g., Schr\"{o}dinger equation, Dirac equation, or decoupled
Maxwell's equations),%
\begin{equation}
{\mathbf{H}}\left( \mathbf{k}\right) \cdot \mathbf{w}_{n}(%
\boldsymbol{\mathbf{k}})=\omega _{n}(\boldsymbol{\mathbf{k}})\mathbf{w}_{n}(%
\boldsymbol{\mathbf{k}}) \label{eigen2}
\end{equation}%
where $\mathbf{w}_{n}\left( \mathbf{k}\right) $ are eigenfunctions and $%
{\mathbf{H}}\left( \mathbf{k}\right) $ is a Hermitian operator
(explicitly defined for the electromagnetic case later) under the inner product $%
\left\langle \mathbf{w}_{m}|\mathbf{w}_{n}\right\rangle =\mathbf{w}%
_{n}^{\dagger }(\boldsymbol{\mathbf{k}})\cdot \mathbf{w}_{m}(\boldsymbol{%
\mathbf{k}})$, with the normalization condition $\left\langle \mathbf{w}_{n}|%
\mathbf{w}_{n}\right\rangle =1$. 

It is crucially important to note that (\ref{eigen2}) defines the eigenmodes only up to a phase
factor, which can depend on the momentum and mode index. We suppose that the eigenmode $\mathbf{w}_{n}(\mathbf{k})$ is initially at
some momentum $\mathbf{k}_{i}$, and evolves along a path in momentum space
such that the mode ends up at some final momentum $\mathbf{k}_{f}$. If we
suppose that the traversed path is closed (the importance of a closed path is discussed below), then $\mathbf{k}_{i}=\mathbf{k}%
_{f}$. Then, under adiabatic evolution of the eigenmode in the momentum domain, and assuming well-defined and single-valued eigenmodes, we have the
boundary condition $\mathbf{w}_{n}(\mathbf{k}_{i})=e^{i\gamma _{n}(\mathbf{k}%
)}\mathbf{w}_{n}(\mathbf{k}_{f})$. This leads to the phase accumulation \cite{Berry}-\cite{Hanson1}
\begin{equation}
\gamma _{n}=\oint_{C}d\boldsymbol{\mathrm{k}}\cdot \boldsymbol{\mathrm{A}}%
_{n}(\mathbf{k})  \label{Eq:Berry_phase}
\end{equation}%
where  
\begin{equation}
\boldsymbol{\mathrm{A}}_{n}(\mathbf{k})=i\mathbf{w}_{n}^{\dagger }(\mathbf{k}%
)\cdot \nabla _{\mathbf{k}}\mathbf{w}_{n}({\mathbf{k}})\label{BC}
\end{equation}%
is called the Berry connection, since it connects the eigenmode $\mathbf{w}%
_{n,\mathbf{k}}$ at point $\mathbf{k}$ and at point $\mathbf{k}+d\mathbf{k}$ via the nonlocal nature of the derivative operator. This cumulative phase effect is called the Berry phase \cite{Berry}. This is the momentum-space analog of the Aharonov-Bohm phase $\phi _{n}=\oint_{C}d\boldsymbol{\mathrm{r}}\cdot \boldsymbol{\mathrm{A}^{\textrm{(m)}}}(\mathbf{r}) $ where $\boldsymbol{\mathrm{A}^{\textrm{(m)}}}$ is the (real-space) magnetic vector potential.

It is worthwhile to note that the Berry phase can be divided into (a) a phase difference corresponding to evolution along one ray as the wave propagates in an inhomogeneous medium, or a curved medium, and (b) a phase difference between different rays (k-vectors) in a homogeneous environment \cite{Hanson1}-\cite{Bliokh}. An example of the latter case is provided below. The change in polarization resulting from propagation of the $\mathrm{TE}_{11}$ mode along a curved circular waveguide provides a simple case of the former \cite{Hanson1}, similar to the case of the helically-wound optical fiber considered in \cite{Chiao}-\cite{TC1986}), and is related to parallel-transport of a vector on a curved surface. 

\section{Gauge Considerations}

Multiplication of the eigenfunction $\mathbf{w}_{n}$ by a
phase factor represents a gauge transformation of
the Berry connection,
\begin{align}
& \boldsymbol{\mathrm{A}}_{n}^{\prime }=i\mathbf{w}_{n}^{\dagger }(\mathbf{k}%
)e^{-i\mathrm{\xi }(\mathbf{k})}\cdot \nabla _{\mathbf{k}}\left( \mathbf{w}%
_{n}(\mathbf{k})e^{i\mathrm{\xi }(\mathbf{k})}\right)   \nonumber \\
& ~~~~=\boldsymbol{\mathrm{A}}_{n}-\nabla _{\mathbf{k}}\xi (\mathbf{k}),
\label{gauge}
\end{align}%
 where $e^{\mathrm{i\xi (\mathbf{k})}}$ an
arbitrary smooth unitary transformation. Therefore, the Berry connection/vector potential is gauge dependent,
like the electromagnetic vector potential. For an arbitrary path, one can choose a suitable gauge such that accumulation of the extra phase vanishes. However, by considering a closed path and noting that the eigenbasis should be single-valued, then a
change of gauge can only change the Berry phase by integer multiples of $2\pi$ \cite{Hanson1}.

\section{Berry Curvature and Electronic Aspects}

In addition to Berry phase and connection, it turns out that the curl of the Berry connection, 
\begin{equation}
\mathbf{F}(\mathbf{k}%
)=\nabla _{\mathbf{k}}\times \boldsymbol{\mathrm{A}}(\mathbf{k}),
\end{equation}%
the Berry curvature, plays a crucially important role in understanding the effects of momentum-space topology. The Berry curvature plays the role of a momentum-space magnetic flux density, analogous to how the real-space magnetic flux density is obtained
as the (real-space) curl of the magnetic vector potential, $\boldsymbol{\mathrm{B}}\left( \mathbf{r}\right) =\nabla _{\mathbf{r}}\times 
\boldsymbol{\mathrm{A}}^{\left( \text{m}\right) }(\mathbf{r})$.

Although our interest here is in photonic problems, it is informative to
consider the effect of electronic Berry effects on the electronic response of
materials. Electronic Berry effects are also sufficient, but not necessary, to have topological photonic effects. Therefore, consider (\ref{eigen2}) as describing electronic
eigenfunctions (e.g., of Schr\"{o}dinger equation), leading to an electronic Berry phase $\gamma ^{\left( \text{e%
}\right) }$ and connection ${\mathbf{A}^{\left( \text{e}\right) }}\left( 
\mathbf{k}\right) $. Here, we use the semiclassical Boltzmann's equation to
illustrate the effect of electronic Berry connection on the material response.

For an electron wavepacket having momentum $\mathbf{p}=\hslash \mathbf{k}$,
Boltzmann's equation is%
\begin{equation}
\frac{\partial f_{\mathbf{k}}}{\partial t}+\overset{\cdot }{\mathbf{r}}\cdot
\nabla _{\mathbf{r}}f_{\mathbf{k}}+\overset{\cdot }{\mathbf{k}}\cdot \nabla
_{\mathbf{k}}f_{\mathbf{k}}=I_{\text{coll.}},
\end{equation}%
where $I_{\text{coll.}}$ accounts for collisions, $f_{\mathbf{k}}$ is the
Fermi distribution, and the electron's equations of motion are \cite{CN}
\begin{align}
\overset{\cdot }{\mathbf{r}}& =\mathbf{v}_{\varepsilon }-\overset{\cdot }{%
\mathbf{k}}\times \mathbf{F}_{\mathbf{k}}^{\left( \text{e}\right) },
\label{rp} \\
\hslash \overset{\cdot }{\mathbf{k}}& =-e\mathbf{E}-e\overset{\cdot }{%
\mathbf{r}}\times \mathbf{B,}  \nonumber
\end{align}%
where $\mathbf{v}_{\varepsilon }\mathbf{=}\frac{1}{\hslash }\nabla _{\mathbf{%
k}}\varepsilon $ and $\mathbf{F}^{\left( \text{e}\right) }(\mathbf{k}%
)=\nabla _{\mathbf{k}}\times \boldsymbol{\mathrm{A}}^{\left( \text{e}\right)
}(\mathbf{k})$ is the electronic Berry curvature. In light of the dual nature to the two equations (\ref{rp}), the Berry curvature can be viewed as
an effective magnetic field in momentum space. In contrast to the Berry connection, the Berry curvature is clearly
gauge-independent, and provides an extra contribution to the velocity, resulting in a current transverse to the applied field
(i.e., a Hall-type/gyrotropic response). 

Solving (\ref{rp}), the current is%
\begin{equation}
\mathbf{J}=e\int \frac{d^{3}k\ f_{\mathbf{k}}}{\left( 2\pi \right) ^{3}}%
\left[ \mathbf{v}_{\varepsilon }+e\hslash ^{-1}\mathbf{E}\times \mathbf{F}_{%
\mathbf{k}}^{\left( \text{e}\right) }+e\hslash ^{-1}\left( \mathbf{F}_{%
\mathbf{k}}^{\left( \text{e}\right) }\cdot \mathbf{v}_{\varepsilon }\right) 
\mathbf{B}\right] ,  \label{J}
\end{equation}
where the summation over all bands is implied. The
first term in (\ref{J}) leads to the usual current and conductivity response, including, in the presence
of a magnetic field, the gyrotropic (Hall-like) response. Of the other two terms in (\ref{J}), the second term $%
\mathbf{E}\times \mathbf{F}_{\mathbf{k}}^{\left( \text{e}\right) }$ is most
important for our purposes, and also leads to a gyrotropic (Hall-like)
tensor response, purely due to (electronic) Berry curvature effects. The Berry curvature term in \ref{J} clearly leads to a current response that is perpendicular to the electric field, due to the anomalous velocity in \ref{rp}. 

\section{Maxwell's Equations as an Eigenvalue Problem \label{SecMax}}

Since source-free Maxwell's equations can be cast as an eigenvalue problem in momentum space, we can define an electromagnetic Berry phase $\gamma ^{\left( \text{em%
}\right) }$, Berry connection ${\mathbf{A}^{\left( \text{em}\right) }}$, and Berry curvature $\mathbf{F}^{\left( \text{em}\right) }$. Similar to the electronic equations of motion \ref{rp}, analogous photon equations of motion (ray equations) for a photonic wave packet have been obtained, \cite{Haldane}
\begin{equation}
\overset{\cdot }{\mathbf{r}} =\mathbf{v}_{g}-\overset{\cdot }{%
\mathbf{k}}\times \mathbf{F}_{\mathbf{k}}^{\left( \text{em}\right) },
\label{rp2} 
\end{equation}%
where $\mathbf{r}$ is the ray trajectory and $\mathbf{v}_{g}$ is the group velocity. The gauge field ${\mathbf{A}^{\left( \text{em}\right) }}$ and it's curvature ${\mathbf{F}^{\left( \text{em}\right) }}$ serve to change the motion of photons in a manner similar to how the electronic Berry connection leads to charged particle motion in helical orbits. 

We consider homogeneous, lossless, bianisotropic material with
frequency-independent dimensionless parameters $\boldsymbol{\varepsilon },~%
\boldsymbol{\mu },~\boldsymbol{\xi },~\boldsymbol{\varsigma }$ representing
permittivity, permeability and magneto-electric coupling tensors, respectively, written as the Hermitian matrix 
\begin{equation}
\boldsymbol{\mathrm{M}}=\left( 
\begin{array}{cc}
\varepsilon _{0}\boldsymbol{\varepsilon } & \frac{1}{c}\boldsymbol{\xi } \\ 
\frac{1}{c}\boldsymbol{\varsigma } & \mu _{0}\boldsymbol{\mu }%
\end{array}%
\right).  \label{matM}
\end{equation}
Source-free Maxwell's equations can be written as a standard Hernitian eigenvalue problem (\ref{eigen2}) 
\cite{Hanson1}, \cite{Mario1}, where ${\mathbf{H}}\left( \mathbf{k}\right) =%
\boldsymbol{\mathbf{M}}^{-1/2}\ \boldsymbol{\mathbf{N(\mathbf{k})~M}}^{-1/2}$ and $\mathbf{w}_{n}(\boldsymbol{\mathbf{k}})=%
\boldsymbol{\mathbf{M}}^{1/2}\cdot \left( \mathbf{E}\text{ }\mathbf{H}%
\right) ^{\text{T}}$, where 
\begin{equation}
\boldsymbol{{\mathrm{N}}}=\left( 
\begin{array}{cc}
0 & \boldsymbol{\mathbf{k}}\times \boldsymbol{\mathrm{I}}_{3\times 3} \\ 
-\boldsymbol{\mathbf{k}}\times \boldsymbol{\mathrm{I}}_{3\times 3} & 0%
\end{array}%
\right).  \label{matN}
\end{equation}%
The electromagnetic eigenfields are of the form $\boldsymbol{w}_{n}(\mathbf{r})=%
\boldsymbol{w}_{n}(\boldsymbol{\mathbf{k}})e^{i\mathbf{k}\cdot \mathbf{r}}$,
where the solution of (\ref{eigen2}), $\boldsymbol{w}_{n}\left( \mathbf{k%
}\right) $, is the envelope of the fields and is independent of position.

\section{Maxwell's Equations Are All That One Needs}
It is worthwhile to point out that the consideration of Berry properties is very helpful in both designing photonic systems and understanding the behavior of such systems. However, Maxwell's equations encompass all relevant phenomena pertinent to Berry phase, connection, and curvature, and there are no additional terms that need to be added to either Maxwell's equations or the constitutive parameters to model topological effects. That is, from a photonic standpoint, the microscopic origin of the material response six-tensor (\ref{matM}) is irrelevant. Simulation codes that solve Maxwell's equations with the usual constitutive parameters inherently model topological effects. 

\section{Symmetry Considerations}

In systems that respect TR symmetry, $\mathbf{F}%
\left( \mathbf{k}\right) =-\mathbf{F}\left( -\mathbf{k}\right) $ and in
systems that respect inversion (parity) symmetry, such as
centrosymmetric systems, $\mathbf{F}\left( \mathbf{k}\right) =\mathbf{F}%
\left( -\mathbf{k}\right) $. In the case of both TR and inversion symmetry, $\mathbf{%
F}\left( \mathbf{k}\right) =\mathbf{0}$, such that breaking one symmetry can result in $\mathbf{F}\left( \mathbf{k}\right) \neq 
\mathbf{0}$. As we see next, having nonzero Berry curvature is essential for some of the most important topological effects in applied electromagnetics. 

\section{Gap Chern Number and TP Edge Modes}

It should be strongly emphasized that the topological aspects mentioned so far concern the behavior of the bulk (infinite, homogeneous) material. Being an integer, $\mathcal{C}_{n}$ must
remain invariant under continuous transformations of momentum space, and can only change (discretely) when, say, an energy bandgap is closed or opened (akin to removing or opening a hole in a geometrical surface, changing its genus). For the structures discussed here, in order to have TP edge modes it is required that  Berry curvature be nonzero, which requires rotational Berry connection.  In general, Berry phase,
connection, and curvature are geometric phenomena that, however, relate to topology, while only the Chern number is a topological invariant (although the Berry phase can be topological in 1D or 2D systems).

In addition to the link between Chern number \ref{CN} and the Gauss-Bonnet theorem \ref{GB}, from an electromagnetic perspective, the Chern number can be interpreted in
an intuitive way from elementary electromagnetics. Gauss's law relates the
total flux over a closed surface $S$ to the total charge within the surface, 
\begin{equation}
\oint_{S}\varepsilon _{0}\boldsymbol{\mathrm{E}}\left( \mathbf{r}\right)
\cdot d\boldsymbol{\mathrm{S}}=Q^{T}=mq,
\end{equation}%
assuming $m$ monopole charges each having charge $q$. Therefore, the normalized flux integral $(1/q)\oint_{S}\varepsilon _{0}\boldsymbol{\mathrm{E}}\left( \mathbf{r}\right)
\cdot d\boldsymbol{\mathrm{S}}=m$ is an integer indicating the number of charge monopoles within the surface. 

The momentum space analog of (\ref{CN}) indicates the number of ``Berry
monopoles" within the surface. Berry monopoles are the momentum-space analog
of a magnetic monopole, and serve as a source/sink of Berry curvature, just
as the electric charge monopole serves as a source/sink of electric
field. If the Chern number is $n$, the net number of Berry monopoles (whose
charges do not cancel each other) is $n$.

Perhaps the most important consequence of having non-trivial topological properties is what happens when the material is truncated, forming a surface with another material. Assume that we have two adjoining materials, at least one of which is topological, each sharing a common bandgap (or the second material is an opaque medium). One can define a type of cumulative Chern number by summing over all bands below the gap, $\mathcal{C}_{%
\text{gap}}=\sum_{n<n_\textrm{gap}}\mathcal{C}_{n}$. It turns out that the difference between the gap Chern numbers of
the two materials ($\mathcal{C}_{\text{gap},\Delta }=\mathcal{C}_{2}-%
\mathcal{C}_{1}$) gives the number of unidirectional surface states that will propagate in the gap (this correspondence is known as the \emph{%
bulk-edge correspondence principle}). It has recently been shown that $\mathcal{C}_{\text{gap}}$ can be computed in terms of the material Green function \cite{GF}, as an alternative to working with the modal functions. Importantly, the unidirectional surface states are topologically-protected, and propagate with no backscattering. The absence of backscattering is insensitive to any surface perturbations, as it is related to the topology of the momentum surface of the bulk energy bands. Additionally, by operating within the bandgap of the two materials, diffraction into the bulk at surface perturbations is also unallowed.

\begin{figure}[ht]
    \begin{center}
        \noindent  \includegraphics[width=3.5in]{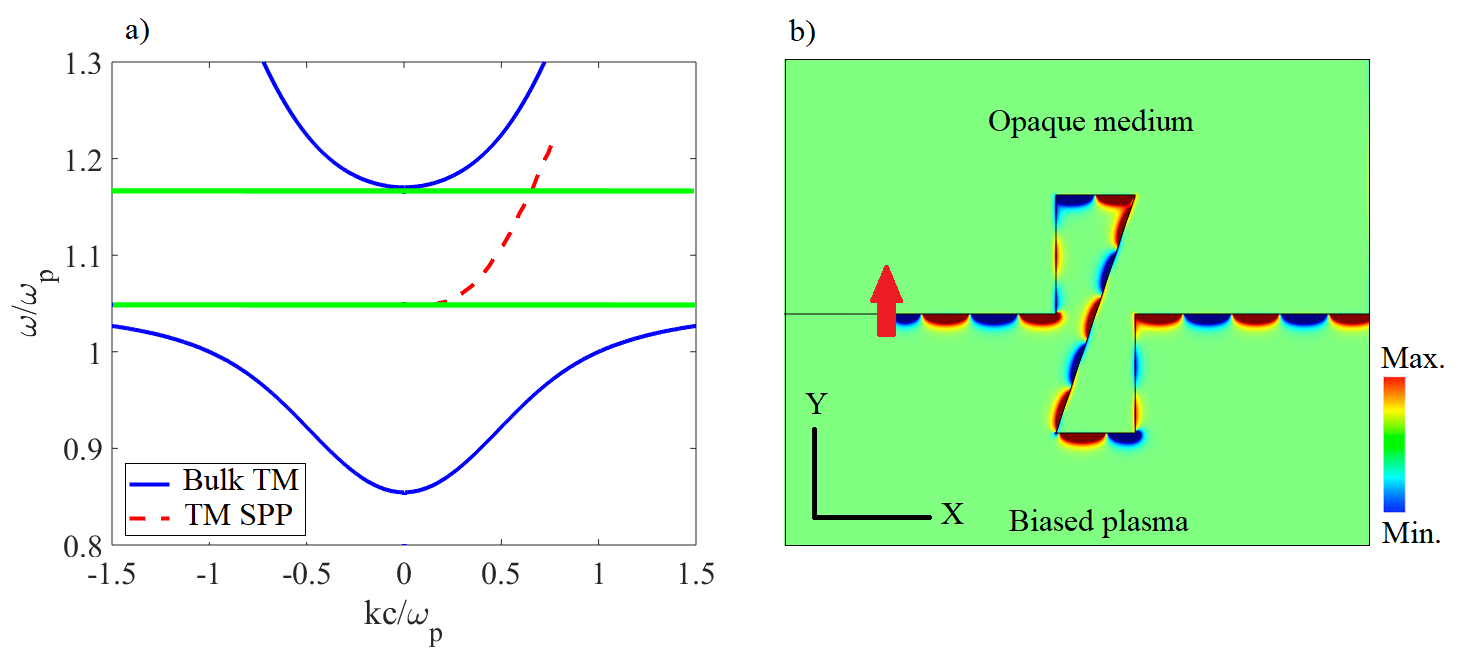}
    \end{center}
    \caption{a. Dispersion behavior of bulk (blue) and SPP (dashed red) modes for the interface between a perfect conductor and a gyrotropic plasma, $\omega_c/\omega_p=0.31$. b. SPP excited by a vertical 2D source at the interface between a biased plasma having $\omega=1.1\omega_p$, and $\omega/\omega_c = 3.5$ and an opaque material having $\varepsilon_s=-2$.}
    \label{NL}
\end{figure}

As an example, albeit not a magnet-less one, Fig. \ref{NL}a shows the bulk dispersion behavior for the transverse-magnetic modes of a gyrotropic plasma ($\omega_p$ is the plasma frequency and $\omega_c$ the cyclotron frequency; the permittivity tensor is defined in \cite{Hanson1}). The gap Chern number is 1, and the TP gap-crossing edge mode (surface-plasmon polariton) is shown as the dashed curve. Fig. \ref{NL}b shows the electric field along the perturbed interface between an opaque medium and the plasma (the size of the perturbation is on the order of a wavelength). It is clear that the edge mode propagates around the electrically-large discontinuity without reflection or diffraction.

Topologically-protected edge modes have been the subject of many studies, and have been experimentally verified \cite{Joannopoulos}-\cite{Rev01}. It can again be emphasized that, given a nonreciprocal material, the occurrence of unidirectional surface propagation is not unexpected, and often occurs even when bulk modes have reciprocal dispersion. However, appreciating the role of topology can be used to, e.g. engineer structures with given Chern numbers \cite{Solja4}. 

\section{Conclusion}

The role of momentum-space topology in nonreciprocal environments has been reviewed. It has been shown that topological concepts can be used to understand and predict novel electromagnetic phenomena associated with nonreciprocity, and particular emphasis has been placed on achieving topologically-protected backscattering- and diffraction-immune edge states at the interface of a topological medium.

\end{document}